# UML Sequence Diagram: An Alternative Model

Sabah Al-Fedaghi
Computer Engineering Department
Kuwait University
Kuwait
sabah.alfedaghi@ku.edu.kw, salfedaghi@yahoo.com

*Abstract*—The UML sequence diagram is the second most common UML diagram that represents how objects interact and exchange messages over time. Sequence diagrams show how events or activities in a use case are mapped into operations of object classes in the class diagram. The general acceptance of sequence diagrams can be attributed to their relatively intuitive nature and ability to describe partial behaviors (as opposed to such diagrams as state charts). However, studies have shown that over 80% of graduating students were unable to create a software design or even a partial design, and many students had no idea how sequence diagrams were constrained by other models. Many students exhibited difficulties in identifying valid interacting objects and constructing messages with appropriate arguments. Additionally, according to authorities, even though many different semantics have been proposed for sequence diagrams (e.g., translations to state machines), there exists no suitable semantic basis refinement of required sequence diagram behavior because direct style semantics do not precisely capture required sequence diagram behaviors; translations to other formalisms disregard essential features of sequence diagrams such as guard conditions and critical regions. This paper proposes an alternative to sequence diagrams, a generalized model that provides further understanding of sequence diagrams to assimilate them into a new modeling language called thinging machine (TM). The sequence diagram is extended horizontally by removing the superficial vertical-only dimensional limitation of expansion to preserve the logical chronology of events. TM diagramming is spread nonlinearly in terms of actions. Events and their chronology are constructed on a second plane of description that is superimposed on the initial static description. The result is a more refined representation that would simplify the modeling process. This is demonstrated through remodeling sequence diagram cases from the literature.

## I. Introduction

In object-oriented analysis, we identify classes by examining usage scenarios, where classes are determined through nouns or noun phrases [1, 2]. This is followed by analyzing classes with the intent of encapsulation (bundling data and methods) while still keeping data and operations separate. Then, the analysis moves to the task of specifying operations, which define the *behavior* of objects, including the communication that occurs between objects by passing messages to one another. In this phase, requirements might demand examination of how an application behaves as a consequence of external events. A behavioral model indicates how software will respond to external events. According to [2], in general, an *event* occurs whenever the system and actor exchange information; in this case, the event is the fact that information has been exchanged. The creation of the behavioral model necessitates the following steps [2]:
(1) Evaluate all use cases to fully understand the *sequence of interaction* within the system;
(2) Identify events that drive the *interaction sequence* and understand how these events relate to specific objects; and
(3) Create a *sequence* for each use case.

### A. Sequence Diagrams

The UML behavioral representation, the sequence diagram (SD), is the second most common UML diagram that represents how objects interact and exchange messages over time [3, 4]. SDs have been used informally for several decades [5]. The first standardization of SDs came in 1992, and since then, there have been several dialects and variations. SDs show how messages are sent between objects or other instances to perform a task. They are used during the detailed design phase, in which precise interprocess communication must be established according to formal protocols. When testing is performed, the behavior of the system can be described as SDs [5].

Shen [6] observed that the general acceptance of SDs can be attributed to their relatively intuitive nature and ability to describe partial behaviors (as opposed to such diagrams as state charts). SDs play an important role in helping users understand system operation and visualize the interactions among a system's objects. According to [7], "the reason of their success compared to other formalisms like state machines is that they are easy to use and understand." The SD bridges the UML use case model and the object classes specified in the structural model. In UML, diagrams must be tightly integrated to avoid inconsistencies; however, such tight integration is infeasible and often impractical [8]. UML 2 extended SDs with such features as recent enrichment and new symbols that allowed programmers to indicate additional procedural details [3, 6].

SDs are used for different purposes, such as showing the flows of method calls inside a program or giving a partial specification of interactions in a distributed system [9]. SDs are utilized to automate test case generation [10]. SDs show "how events or activities in a use case are mapped into operations of object classes in the class diagram. Events are basic behavioral constructs of SDs that can be combined to form larger behavioral constructs called fragments" [11].

*B. Problems*

Many different semantics have been proposed for SDs for various purposes (e.g., in terms of translations to state machines) [9, 11]. In short, there does not exist a suitable semantics-based refinement of required SD behavior because direct style semantics do not precisely capture required SD behaviors, and translations to other formalisms disregard essential features of SDs such as guard conditions and critical regions [11]. It is not easy to select suitable semantics: the various formal semantics for SDs handle even the most basic diagrams quite differently [9]. Thus, there is a need for a refinement of SD behaviors that clarifies the issue of the relationship of the static and dynamic features.

Modeling complex interaction behaviors relies on a good understanding of SDs [11]. However, SDs often pose the greatest difficulties among novices learning modeling [11, 12]. Modeling SDs overwhelms some learners, as it involves a large number of interacting items that must be handled concurrently [13].

Over 80% of graduating students are unable to create a software design or even a partial design [14]. According to [14], "many students had no idea how SDs were constrained by other models. Many exhibited difficulties in identifying valid interacting objects and constructing messages with appropriate arguments. Though students understood the role of objects, messages and arguments individually, they were daunted when considering all constraints imposed by other models, concurrently."

*C. Generalizing SDs*

This paper proposes an alternative generalized model to SDs that provides further understanding of SDs, assimilating them into a new modeling language called a thinging machine (TM). The SD is extended horizontally, removing the superficial dimensional limitation of vertical-only expansion to preserve the logical chronology of events. TM diagramming is spread nonlinearly in terms of actions. Events and their chronology are constructed on a second plane of description that is superimposed on the initial static description. The result is a more refined representation that simplifies the modeling process. This is demonstrated by remodeling SD cases from the literature.

For the sake of having a self-contained paper, the next section introduces TM as our main tool in scrutinizing SDs. Sections 3 and 4 present case studies that contrast SDs with the proposed TM modeling.

## II. TM Modeling

The TM model articulates the ontology of the world in terms of an entity that is simultaneously a *thing* and a *machine*, called a *thimac* [15-25]. A thimac is like a double-sided coin. One side of the coin exhibits the characterizations assumed by the thimac, whereas on the other side, operational processes emerge that provide dynamics. A thing is subjected to doing, and a machine does.

Thimacs are a source for generic constructs that can be applied in conceptual modeling to describe structure and behavior as a world of systems (thimacs). The generic actions in the machine (see Fig. 1) can be described as follows:

**Arrive:** A thing moves to a machine.

**Accept:** A thing enters the machine. For simplification, we assume that all arriving things are accepted; hence, we can combine the *arrive* and *accept* stages into one stage: the **receive** stage.

**Release:** A thing is ready for transfer outside the machine.

**Process:** A thing is changed, but no new thing results.

**Create:** A new thing is born in the machine.

**Transfer:** A thing is input into or output from a machine.

Additionally, the TM model includes storage and triggering (denoted by a dashed arrow in this study's figures), which initiates a flow from one machine to another. Multiple machines can interact with each other through movement of things or triggering. Triggering is a transformation from one series of movements to another.

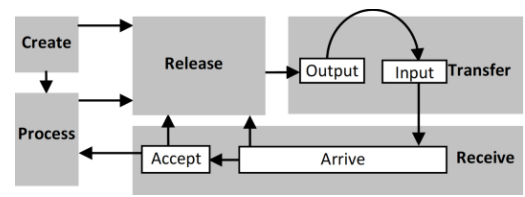

Fig. 1. Thinging machine.

## III. ATM Example

Consider the simple SD for withdrawing cash from an ATM shown in Fig. 2 [26].

*A. Static TM Model*

Fig. 3 shows the corresponding TM model. The figure illustrates the following:
1. The user inserts his/her card (1), which flows (2) to the ATM to be processed (3). The ATM extracts the card's number (4), which is sent to the bank system (5).

Note: In the original SD, the message from the ATM to the bank is "verify card". Here, the message does not state what is sent.

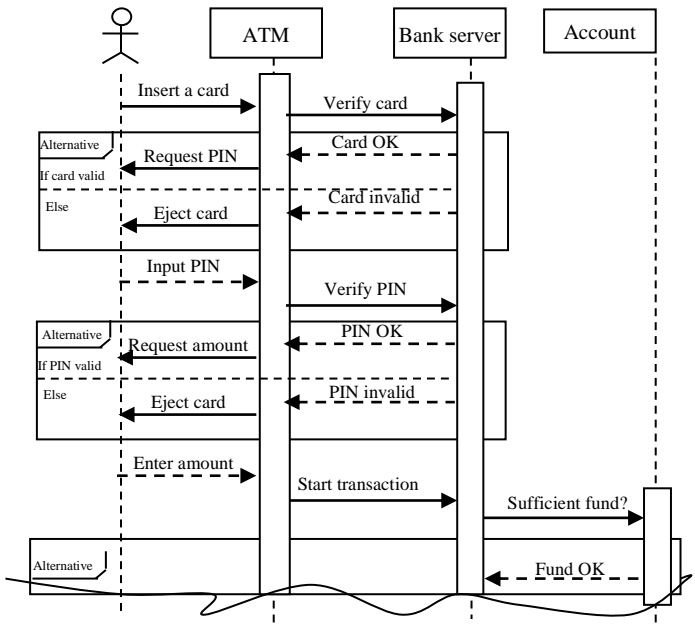

Fig. 2. Sample sequence diagram. (Redrawn, incomplete from [26])

What is sent can be misunderstood as the physical card, which is called a "message". The correct understanding is that the ATM sends the embedded card *number* to be verified. This ambiguity about what is sent can be eliminated by explicitly specifying the process of extracting the number from the physical card. In a TM, the arrow from the user to the ATM is a physical card, and the arrow from the ATM to the bank is a number; these two arrows never cross.

2. The card number is processed (6) in the bank system (not in the bank itself) as valid (7) or invalid (8). If the card number is valid, then an OK message (9) is sent to the ATM. The ATM processes the message (10) and triggers the construction of a request for a PIN (11) that flows to the user (12). If the card number is invalid, then the bank system constructs a "not OK" message (13), which flows to the ATM to trigger the ejection (14) of the card.

Note: This TM part corresponds to *Alternatives* in the SD. The flows of different kinds of things (card, card number, OK/not OK, PIN messages) are separated by triggering.

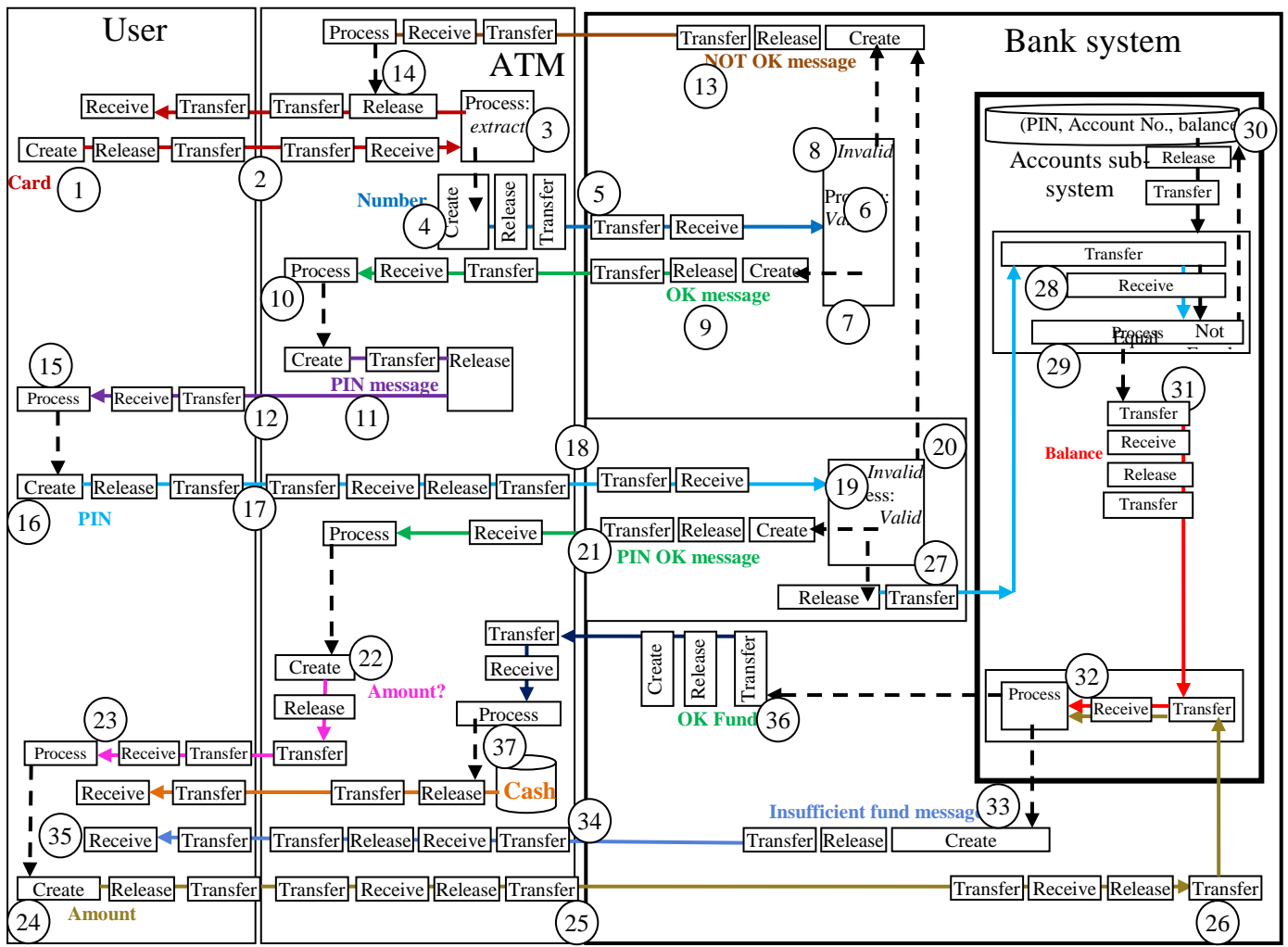

Fig. 3. Static model of the ATM.

3. The user processes (reads; 15) the message to trigger the creation of the PIN (16) that flows to the ATM (17), which sends it to the bank system (18). In the bank system, the PIN is processed (19). If the PIN is invalid (20), a not-OK message (13) is sent to the ATM, causing the ejection of the card (14). If the PIN is valid (21), an OK message (22) is sent to the ATM. The ATM constructs a message asking for the amount, which is sent to the user (23).

Note: The above description corresponds to the second *Alternative* in the SD. The TM model does not repeat the process of ejecting the card when a PIN is invalid (13 and 14).

4. The user enters the amount (24), which is received by the ATM and sent to the bank system (25), which in turn sends it to the account subsystem (26).

Note: The account system is a subsystem of the bank system. It is important to know that the bank system stored the previously given, say, PIN because the account subsystem needs such information to allocate the amount and decide whether the account is sufficient. This is performed when the PIN is valid (27). The SD skips over these parts of the scenario, leaving the model either incomplete or disconnected.

5. The account subsystem receives the PIN (28) and searches for the corresponding account or amount in its database (i.e., PIN, account number, balance). Accordingly, the corresponding balance is extracted (31) and compared (32) with the requested amount. If the funds (balance) are insufficient, then an insufficient funds message (33) is sent to the ATM, which in turn sends it to the user (34 and 35). If funds are sufficient, then an "OK fund" message (36) is sent to the ATM. The ATM releases the corresponding amount of cash to the user (37).

### B. Simplification (if needed)

Of course, sequencers will complain that the TM model is more complex than their SD. There is no argument against the model being simple, however—the model must include as many important situations as is practical. Simplicity should not reduce reasonable completeness. In the ATM example, it is reasonable and important to expect that what is sent to the bank would be the number embedded in the card. It is also reasonable and important to indicate that the amount message (sufficient/insufficient) sent by the account system involves searching for a balance in the account database. Such omissions are common in SDs. The additions supplied by the TM model to the ATM problem reflect more or less relevant expansions to the conception of the problem. Fig. 3 can be simplified by assuming that the arrow direction indicates the direction of flow; thus, the transfer, release, and receive steps can be eliminated, resulting in Fig. 4.

### C. Time and Behavior

Additionally, time in SDs is said to be represented along the vertical dimension. However, this confuses time with logical order. Suppose that the numbers 1, 2, and 3 are listed vertically in order. This does not mean that 1 *happens* at time 1, 2 happens at time 2, and 3 happens at time 3. The relationship among 1, 2, and 3 is the logical relationship "less than" or "greater than", and there are no time-related events involved. To make the relationships among 1, 2, and 3 into events, we must introduce the idea of "existence" or "presence" ("creation" in a TM). Suppose that for a certain system, at time zero, 3 is created (born in the context of the system), and then 1 is created. Then, we can have the events (time zero, 3) and (time zero + 1, 1). In this case, we can say that 3 happened before 1, regardless of the logical relationship between 1 and 3 (e.g., less than). Of course, with this understanding, we can simplify the events and write them as *3 then 1* (without mentioning times explicitly).

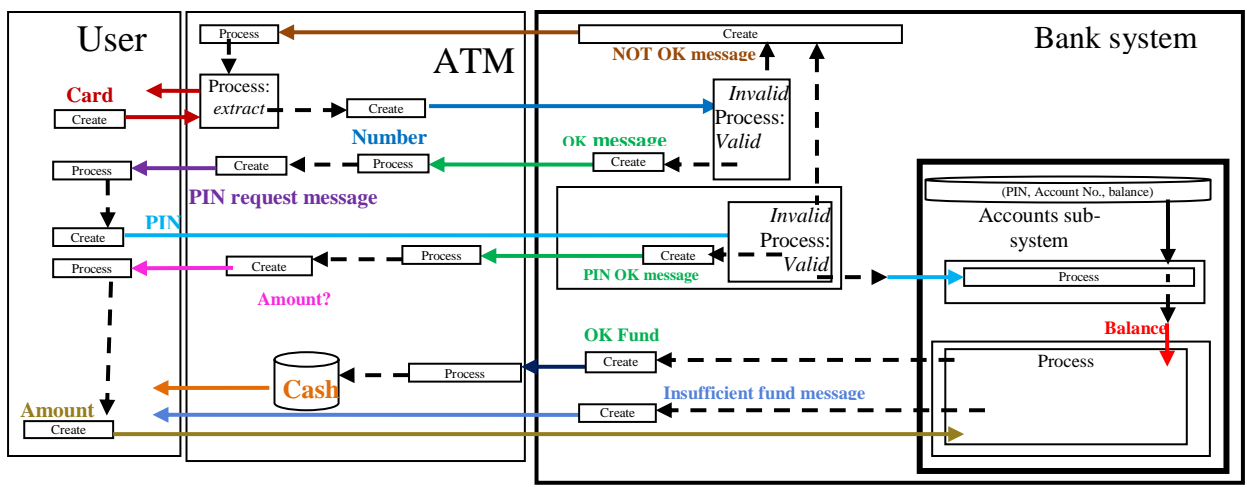

Fig. 4. Simplification of static ATM model.

Another issue is that the SD forces non-logical relationships. For example, the card and the PIN can be entered in any order or concurrently (e.g., to speed up the transaction). Because these are independent and multiple methods of verification, their order and execution are logically immaterial. Such a situation can be observed clearly when websites ask for an account and password simultaneously: when one of them is wrong, the system does not tell you which one. Incorporating this multi-verification case requires extending the SD. Still, the basic conceptual problems in the SD are claiming that the vertical order is time, failing to define what time is, and failing to define what an event is.

A model exists in time as much as space, and it must situate itself in the temporal. The TM model fuses space and time into a single dynamic model of events. TM modeling involves two planes of modeling: staticity and dynamics. The static model involves spatiality (containerization) and actionality (generic actions). Spatiality involves recognizing the thimac areas that partition the model, taking into account the connections (flows and triggering) between these areas.

A union of TM spatiality/actionality with time defines Event 5. The event blends such a spatiality/actionality thimac with time. Actuality here means the five generic actions. In the ATM example, Fig. 5 shows the machine of the event *the user inserts a card that is received by the ATM*. Accordingly, the static model (Fig. 3) is divided into parts; each represents the region of an event, as shown in Fig. 6.

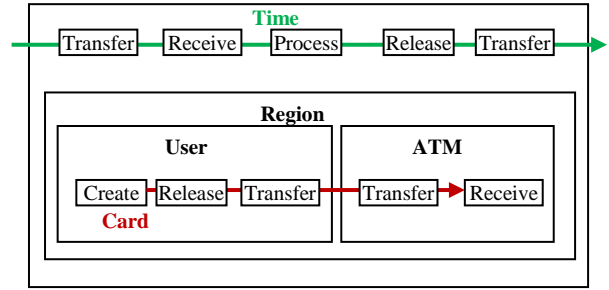

Fig. 5. The event *the user inserts a card that is received by the ATM*.

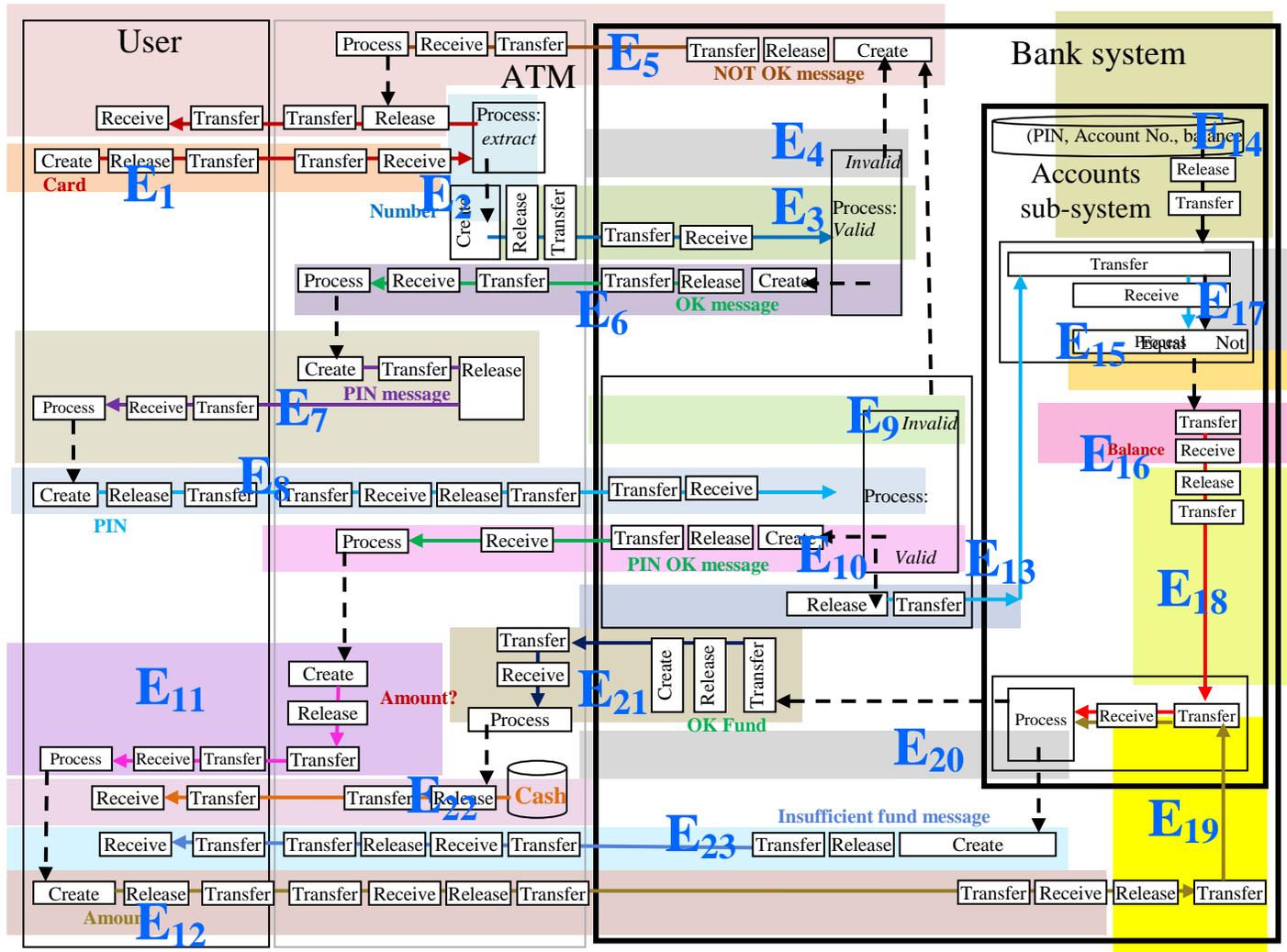

Fig. 6. Dynamic ATM model, regions of events.

The SD does not include the notion of time because an event is made up of the time + region of the event, which includes the boundary and actions. We can specify the events in the ATM model as follows.

Event 1 ($E_1$): The user inserts a card that flows to the ATM.

Event 2 ($E_2$): The ATM processes the card and extracts its number.

Event 3 ($E_3$): The card number is sent to the bank, where it is processed.

Event 4 ($E_4$): The card number is invalid.

Event 5 ($E_5$): A not-OK message is sent to the ATM, and the ATM ejects the card.

Event 6 ($E_6$): The card number is valid; hence, an OK card number message is sent to the ATM.

Event 7 ($E_7$): The ATM requests the PIN.

Event 8 ($E_8$): The user inputs the PIN, which flows to the ATM, which sends it to the bank.

Event 9 ($E_9$): The PIN is invalid.

Event 10 ($E_{10}$): The PIN is valid; hence, an OK PIN message is sent to the ATM.

Event 11 ($E_{11}$): The ATM requests the amount.

Event 12 ($E_{12}$): The user inputs the amount, which flows to the ATM, which sends it to the bank.

Event 13 ($E_{13}$): The bank system sends the PIN to the database system.

Event 14 ($E_{14}$): The database system retrieves a PIN record.

Event 15 ($E_{15}$): The database system compares the user's PIN with the record's PIN.

Event 16 ($E_{16}$): The two PINs match; hence, the corresponding balance is extracted.

Event 17 ($E_{17}$): The two PINs do not match.

Event 18 ($E_{18}$): The balance flows to be compared with the amount.

Event 19 ($E_{19}$): The amount flows to be compared with the balance.

Event 20 ($E_{20}$): The balance and amount are compared.

Event 21 ($E_{21}$): The amount of funds is OK; hence, a message is sent to the ATM.

Event 22 ($E_{22}$): The ATM disburses cash.

Event 23 ($E_{23}$): Funds are insufficient; hence, a message is sent to the ATM that, in turn, sends it to the user.

Fig. 7 shows the resultant behavioral model of the ATM.

### D. Contrasting the SD with the TM Model

Consider contrasting the two representations:
- The SD with its objects, messages, communication arrows, events, vertical chronology of events, fragments, activities (e.g., insert, verify, eject, request, input, start, enter, etc.) … (See Fig. 2)
- The TM model with its things, machines, five actions, two types of arrows, events, behavior… (See Figs. 3, 6, and 7).

It seems that the TM model is more systematic (coherent whole), with a clear separation of the static and dynamic (e.g., events) aspects of the system. Such a claim will be substantiated further with the modeling case in the next section.

### IV. ROLE OF THE SD IN UML

The SD is a UML interaction diagram that *bridges* the user requirements specified in the use case model and the object classes specified in the structural model [27]. Syn [13] discussed the example shown in Fig. 8 to demonstrate the systems analysis process, by which the class diagram is derived from use cases. According to [13], the figure "shows the important role of the sequence diagram. The sequence diagram bridges the requirements specified in the use case model with operations of object classes in the structural model."

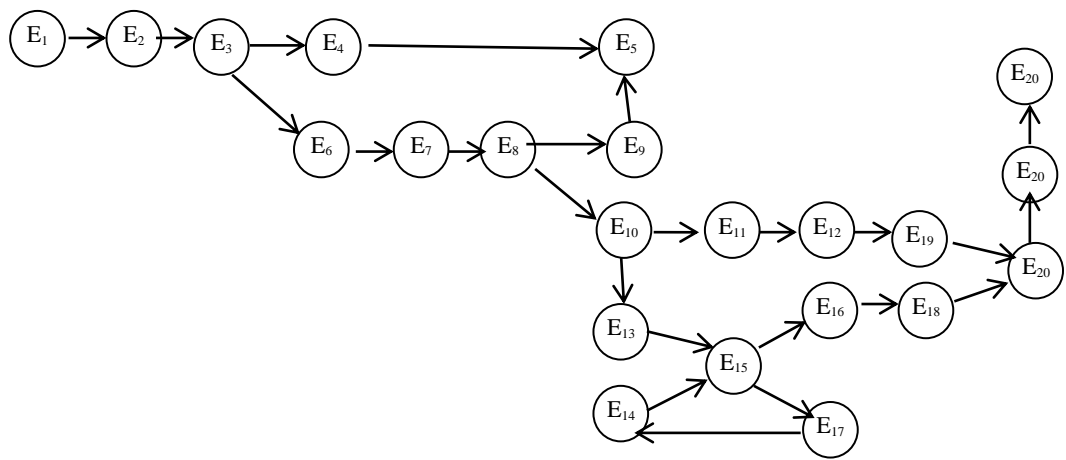

Fig. 7. Behavioral model of the ATM.

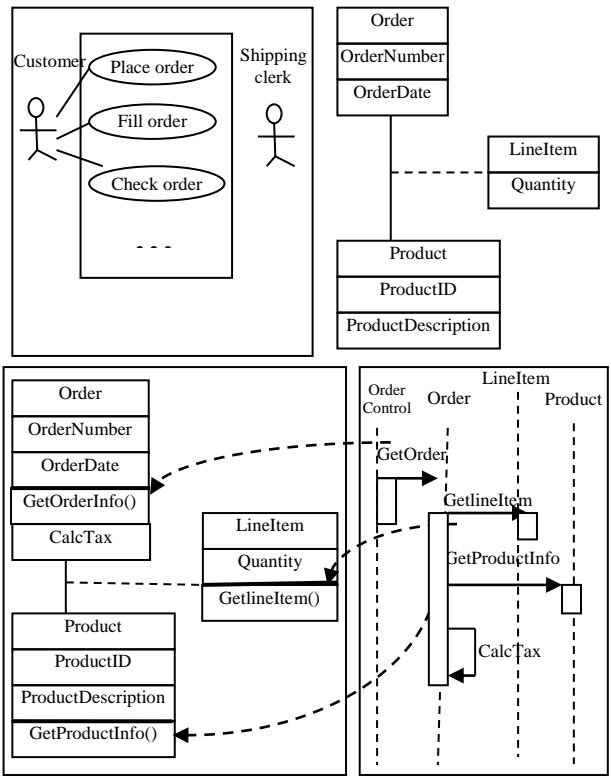

Fig. 8. Sample UML systems analysis process. (Redrawn, incomplete from [13])

*A. Static Model*

Fig. 8 reflects how complicated the integration of the UML diagrams is and the role of the SD in such a process. By contrast, a TM has a single diagram that makes it easy to develop an integrated and consistent specification. Fig. 9 shows the static TM model of Syn's [13] ordering system built according to our understanding of Syn's description with minimum additions required to fill the gaps in the given details.

- In Fig. 9, there are two spheres (machines): the customer (Circle 1) and the company (2). The customer creates an order (3). Note that the order structure (4) is similar to a UML class diagram without the methods.
- The order then flows to the company (5), where it is received (6) and processed (7).
- This processing extracts the product number (8) from the order and sends this number to a module (9) that extracts information about that product.
- Accordingly, a process starts for the record of that product in the product database (10). In such a process, a record is retrieved from the product file (11), processed (12), and the product number is extracted (13). Hence, the two product numbers (from the order and from the system file) are compared (14).
  - If they are not the same (15), a new record is retrieved from the database (16).
  - If they are the same, the product price (17) and description (18) are extracted from the product record.
- From the price (17) and the quantity (19 – extracted from the order), the total price is calculated (20). Tax (21) is also calculated. It is possible to develop the invoice here, but because [13] did not mention such an item, we ignore it.
- Finally, the order (22) and the product description (23) are sent to the inventory warehouse to retrieve the actual product (24) and deliver it to the customer (25).

*B. Dynamic Model*

We continue developing the behavior of Syn's [13] ordering system. First, decomposability is applied to form events. Then, the chronology of events is identified to specify behavior. Fig. 10 shows the decomposition of the static model into regions of events. For simplicity's sake, the time flow is not shown. Accordingly, we have the following events in Syn's [13] ordering system.

Event 1 ($E_1$): A customer creates an order.
Event 2 ($E_2$): The order flows to the company.
Event 3 ($E_3$): The order is processed to extract the product number.
Event 4 ($E_4$): The product number flows to be processed to retrieve the corresponding record from the database.
Event 5 ($E_5$): A record is retrieved from the product database.
Event 6 ($E_6$): The product number is extracted from the retrieved record.
Event 7 ($E_7$): The product number of the retrieved record flows to be compared with the order product number.
Event 8 ($E_8$): The product number in the order is not the same as the product number in the retrieved record.
Event 9 ($E_9$): The product number in the order is the same as the product number in the retrieved record.
Event 10 ($E_{10}$): The product price is extracted.
Event 11 ($E_{11}$): The product description is extracted.
Event 12 ($E_{12}$): The product price flows to a procedure that calculates the total price.
Event 13 ($E_{13}$): The quantity in the order is extracted from the order.
Event 14 ($E_{14}$): The quantity flows to the procedure that calculates the total price.
Event 15 ($E_{15}$): The total price is calculated.
Event 16 ($E_{16}$): The tax is calculated based on the total price.
Event 17 ($E_{17}$): The order flows to the inventory system.
Event 18 ($E_{18}$): The product description flows to the inventory system.
Event 19 ($E_{19}$): The order and the product description are used to retrieve the actual product.
Event 20 ($E_{20}$): The actual product is sent to the customer.

Fig. 11 shows the behavioral model of the ordering system.

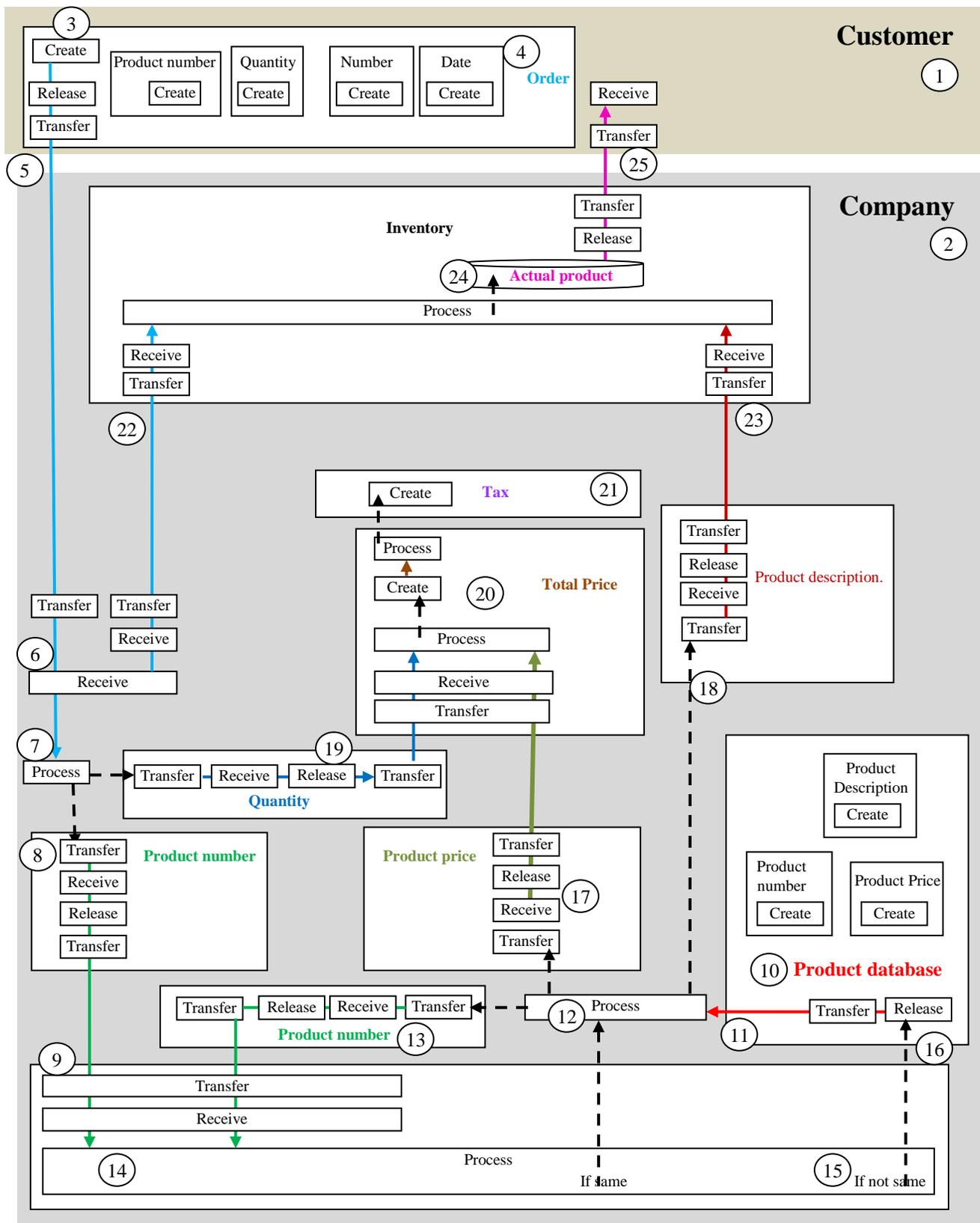

Fig. 9. Static TM model of Syn's [13] ordering system.

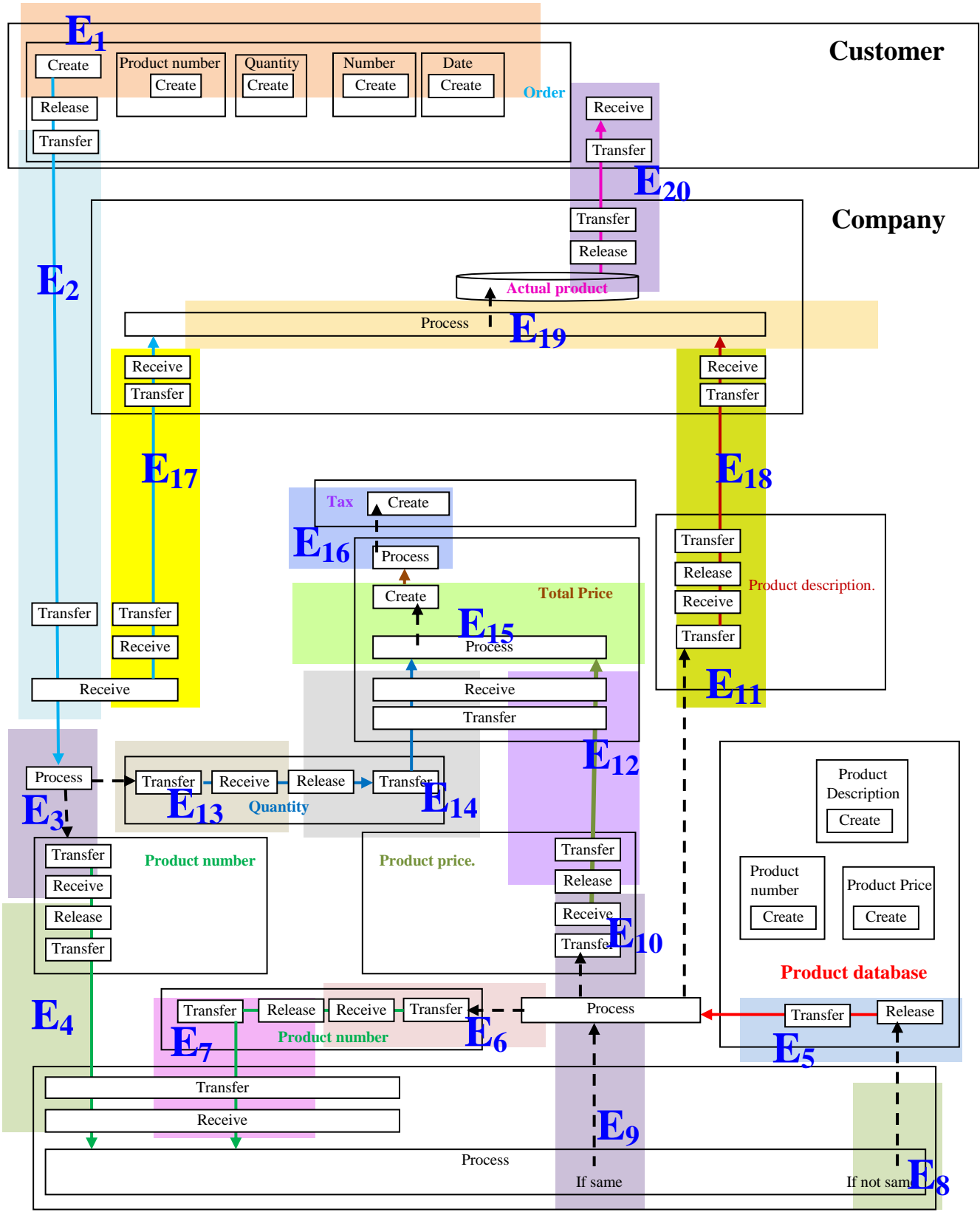

Fig. 10. Decomposition of the static model.

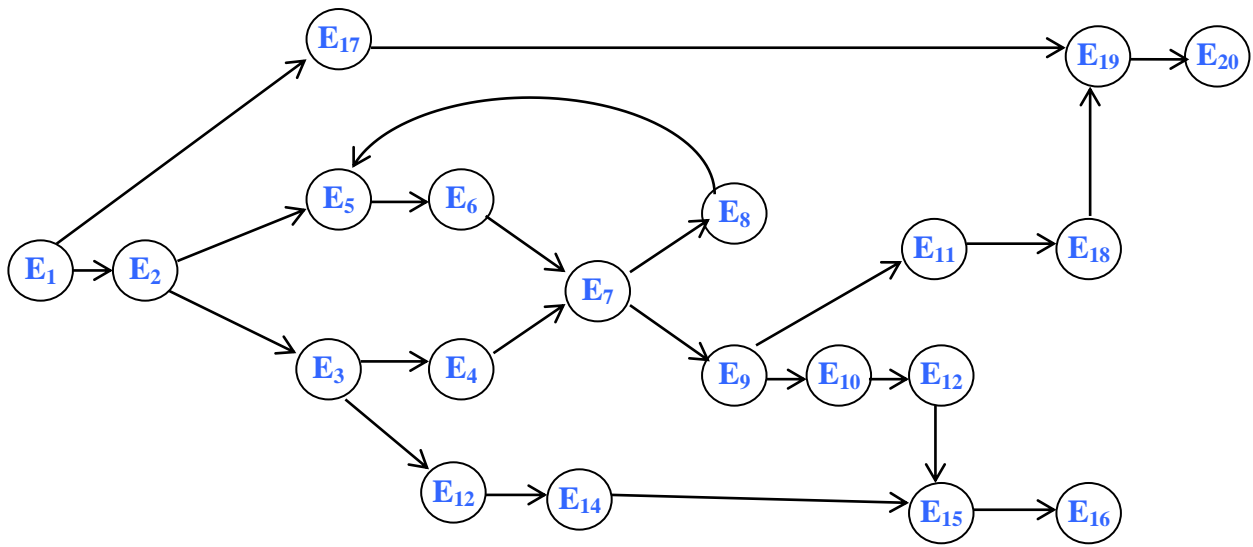

Fig. 11. Behavioral model of the ordering system.

## V. CONCLUSION

In this paper, we proposed an alternative generalized model to the UML SD. This was motivated by difficulties mentioned in the literature regarding developing a suitable semantics-based refinement of required SD behavior. Additionally, studies have shown that graduating students were unable to create software designs, and many students exhibited difficulties in identifying valid interacting objects and constructing messages with appropriate arguments.

The proposed modeling methodology, called TM, extends the modeling process horizontally by removing the superficial dimensional limitation of vertical-only expansion to preserve the logical chronology of events. TM diagramming is spread nonlinearly in terms of actions. Events and their chronology are constructed on a second plane of description that is superimposed on the initial static description. The result is a more refined representation that simplifies the modeling process. We demonstrated this by remodeling SD cases from the literature. TM modeling can be applied to all systems that incorporate SDs. We claim that the results would be a clearer description with better semantics based on the notion of TM actions and events.

The TM model is more systematic (a coherent whole), with clearer separation of the static and dynamic aspects of the system than SD modeling. Our claim is substantiated by the two remodeled study cases above. Accordingly, it seems that adopting this new approach requires studying how to integrate the TM model into the UML diagramming apparatus. Such an issue is a research topic for future work.